\documentclass[12pt]{article}
\usepackage{amsmath,amsthm,amsfonts,amssymb}
\usepackage{graphicx}
\usepackage{enumerate}
\usepackage{natbib}
\usepackage{url} 
\usepackage{array}
\usepackage{amscd}
\usepackage{color}
\usepackage{mathrsfs}
\usepackage{latexsym}
\usepackage{bm,bbm}
\usepackage{extarrows}
\usepackage[colorlinks=true, linkcolor=blue, citecolor=blue]{hyperref}
\usepackage{algorithm}
\usepackage{algpseudocode}

\usepackage{colortbl,xcolor,multirow}
\newcolumntype{f}{>{\columncolor{lightgray}}l}
\newcolumntype{h}{>{\columncolor{lightgray}}r}

\newcommand{\blind}{1}

\newtheorem{definition}{Definition}
\newtheorem{theorem}{Theorem}

\newcommand{\gbf}[1]{\boldsymbol{#1}}
\newcommand{\E}{E}

\begin{document}

\def\spacingset#1{\renewcommand{\baselinestretch}%
{#1}\small\normalsize} \spacingset{1}


\if1\blind
{
  \title{\bf A note on parameter orthogonality for multi-parameter distributions}
  \author{Changle Shen, Dong Li\thanks{
    Li's work is supported in part by the NSFC (no.72471127).}\hspace{.2cm}\\
    Department of Statistics and Data Science, Tsinghua University, \\Beijing 100084, China.\\
    and\\
    Howell Tong \\
    Department of Statistics and Data Science, Tsinghua University, \\
	Paula and Gregory Chow Institute for Studies in Economics, \\
    Xiamen University \\
	Department of Statistics, London School of Economics}
  \maketitle
} \fi

\if0\blind
{
  \bigskip
  \bigskip
  \bigskip
  \begin{center}
    {\LARGE\bf A note on parameter orthogonality for multi-parameter distributions}
\end{center}
  \medskip
} \fi

\bigskip
\begin{abstract}
	This note addresses issues raised by Cox and Reid in their seminal paper in 1987 regarding parameter orthogonality in statistical inference. We extend the orthogonality condition to cases with multiple parameters of interest and demonstrate its existence at a global level for some generally important distributions, despite previously expressed pessimism by them. Numerical results with the location-scale $t$-distribution reveal substantial gains in estimation accuracy and savings in computation time, thanks to the existence. We next show that the local parameter orthogonality can lead to efficient computational algorithms with the celebrated Whittle algorithm for multivariate autoregressive modeling as a showcase.
\end{abstract}

\noindent%
{\it Keywords:} Parameter orthogonality, location-scale $t$-distribution, multivariate autoregressive model, Whittle algorithm. 
\vfill

\newpage
\spacingset{1.9} 

\section{Introduction}
\label{sec:intro}

In their seminal paper, \cite{Cox1987Parameter} initiated an approach to parameter orthogonality which is defined with respect to the Fisher information matrix as described in their Section~\ref{sec:LPO}. They constructed orthogonality of one single parameter of interest $\psi$ to a new set of nuisance parameters $\gbf{\lambda} = (\lambda_1, \ldots, \lambda_q)^\top$ by reparameterization, with the solution to the
partial differential equations~\eqref{eq:CoxReid} 
\begin{equation}
    \label{eq:CoxReid}
    \sum_{r=1}^{q} I_{\gamma_r, \gamma_s} \frac{\partial \gamma_r}{\partial \psi} = - I_{\psi, \gamma_s}, \quad s = 1, \ldots, q,
\end{equation}
playing the pivotal role, where $\gamma_1, \ldots, \gamma_q$ are the original nuisance parameters and $I_{\cdot, \cdot}$ is the corresponding Fisher information. They showed that the reparameterization can lead to substantial gains in efficiency in statistical inference, in the sense that the impact of the estimation error in $\hat{\gbf{\lambda}}$ on $\hat{\psi}_{\hat{\gbf{\lambda}}}$, the maximum likelihood estimate (MLE) of $\psi$ given $\hat{\gbf{\lambda}}$, is asymptotically negligible. They also proposed a likelihood ratio statistic constructed from the conditional distribution given the MLEs for $\gbf{\lambda}$ and justified its preference over the usual profile likelihood function.

\cite{Cox1987Parameter} leaves some open issues unanswered in Section 6 of their paper. For example, item (viii) states `\textit{Can the discussion usefully be extended to vector parameters of interest, where in general only local orthogonality is possible?}\,'; item (ix) states `\textit{How should the differential equations determining $\gbf{\lambda}$ be handled when simple explicit solution is not feasible? What further conditions can usefully be imposed on $\gbf{\lambda}$ in general?}' As far as we know, their questions have not attracted much attention. In particular, Equation~\eqref{eq:CoxReid} has not been explicitly extended to address the above two issues. In this note we aim to fill the gap. We then explored some important distributions for which global parameter orthogonality holds. In particular, for the location-scale $t$-distribution, estimation accuracy and computation time of MLE procedure could be improved if carried out in the orthogonal parameterization. After achieving this, we add one more item to the list of benefits exhibited by~\cite{Cox1987Parameter}, namely that the extension can, in suitable settings, bring about substantial gains in computational efficiency in Statistics, Signal Processing and many other fields. Specifically, we show that it leads to the celebrated~\cite{Whittle1963Fitting} algorithm for a fast inversion of a block Toeplitz matrix central to multivariate autoregressive (AR) modeling. 

\textit{Notations}: Given a zero-mean $m$-dimensional time series $\{\gbf{X}_t\}$, we write $\gbf{\Gamma}_h = \E(\gbf{X}_t\gbf{X}_{t-h}^\top)$ as its lag-$h$ cross-covariance matrix. Let $a_{i}$ represent the $i$-th entry of an $m$-dimensional vector $\gbf{a}$. For an $m$-variate function $f(\gbf{a})$, we use the $m$-dimensional vector $\partial f/ \partial \gbf{a}$ to denote its first-order derivative. For an $m \times m$ matrix $\gbf{A}$, $A_{(i, j)}$ denotes its $(i, j)$-th entry. We denote by $\gbf{E}_{i, j}$ the $m \times m$ matrix with the $(i, j)$-th entry being 1 and all other entries being 0; or equivalently, $E_{i, j, (k, \ell)} = \delta_{i, k} \delta_{j, \ell}$, where $\delta_{i,j}$ is the Kronecker delta function. The symbol ${}^\top$ denotes the transpose of a vector or a matrix.

\section{Parameter orthogonality for multiple parameters of interest}
\label{sec:LPO}

Following~\citet[p. 207]{Jeffreys1961Probability}, we first give a formal definition of parameter orthogonality at both global and local levels.

\begin{definition}
	Consider a sample of $m$-dimensional random vectors $\gbf{X}_1, \ldots, \gbf{X}_n$ drawn independently from a distribution with $\theta \in \gbf{\Theta} \subset \mathbb{R}^d$ being the $d$-dimensional unknown parameters. We denote by $l(\gbf{\theta}; \gbf{X}_{1:n})$ the log-likelihood function, where $\gbf{X}_{1:n} = (\gbf{X}_1^\top, \ldots, \gbf{X}_n^\top)^\top$ is the $n \times m$ matrix of observations. For a partition of $\gbf{\theta} = (\gbf{\theta}_1^\top, \gbf{\theta}_2^\top)^\top \in \gbf{\Theta}$ with $\gbf{\theta}_1 \in \mathbb{R}^{d_1}$, $\gbf{\theta}_2 \in \mathbb{R}^{d_2}$ and $d_1 + d_2 = d$, $\gbf{\theta}_1$ is defined to be globally orthogonal to $\gbf{\theta}_2$ if the entries of the Fisher information matrix satisfy
    \begin{equation}
        \label{eq:ortho}
        I_{i, j}(\gbf{\theta}) = \frac{1}{n} \E \Big\{\frac{\partial l(\gbf{\theta}; \gbf{X}_{1:n})}{\partial \theta_i} \frac{\partial l(\gbf{\theta}; \gbf{X}_{1:n})}{\partial \theta_j} \Big\} = -\frac{1}{n} \E \Big\{\frac{\partial^2 l(\gbf{\theta}; \gbf{X}_{1:n})}{\partial \theta_i \partial \theta_j} \Big\} = 0
    \end{equation}
    for any $i = 1, \ldots, d_1$ and $j = d_1 + 1, \ldots, d$. If~\eqref{eq:ortho} holds only at some point $\gbf{\theta} = \gbf{\theta}^{(0)}$, then $\gbf{\theta}_1$ and $\gbf{\theta}_2$ are said to be locally orthogonal at $\gbf{\theta}^{(0)}$. 
\end{definition}

We now discuss the case in which $\gbf{\theta}_1$ are the parameters of interest and $\gbf{\theta}_2$ are the nuisance parameters and we wish that, via suitable transformations, they become orthogonal. To avoid confusion, we write $\gbf{\theta}_1 = \gbf{\psi} = (\psi_1, \ldots, \psi_{d_1})^\top$ and $\gbf{\theta}_2 = \gbf{\gamma} = (\gamma_{1}, \ldots, \gamma_{d_2})^\top$. Suppose we wish to reparameterize $(\gbf{\psi}^\top, \gbf{\gamma}^\top)^\top$ to $(\gbf{\psi}^\top, \gbf{\lambda}^\top)^\top$, where $\gbf{\lambda} = (\lambda_1, \ldots, \lambda_{d_2})$, with inverse mapping $\bar{\gbf{\gamma}} = (\bar{\gamma}_1, \ldots, \bar{\gamma}_{d_2})^\top$ satisfying
\begin{equation*}
    \gamma_j = \bar{\gamma}_j(\gbf{\psi}, \gbf{\lambda}), \quad j = 1, \ldots, d_2,
\end{equation*}
such that $\gbf{\psi}$ is orthogonal to $\gbf{\lambda}$. The following theorem gives a sufficient condition for the orthogonality.

\begin{theorem}
    \label{thm:1}
    Denote by $I_{\psi_i, \gamma_s}$ the Fisher information between $\psi_i$ and $\gamma_s$, and $I_{\gamma_r, \gamma_s}$ the Fisher information between $\gamma_r$ and $\gamma_s$ in the $(\gbf{\psi}, \gbf{\gamma})$ parametrization, i.e. 
    \begin{equation*}
        I_{\psi_i, \gamma_s} = - \frac{1}{n} \E \Big\{\frac{\partial^2 l(\gbf{\psi}, \gbf{\gamma}; \gbf{X}_{1:n})}{\partial \psi_i \partial \gamma_s} \Big\}, \quad I_{\gamma_r, \gamma_s} = - \frac{1}{n} \E \Big\{\frac{\partial^2 l(\gbf{\psi}, \gbf{\gamma}; \gbf{X}_{1:n})}{\partial \gamma_r \partial \gamma_s} \Big\}.
    \end{equation*}
    Then the orthogonality condition can be expressed as
    \begin{equation}
        \label{eq:ortho_entry}
        \sum_{r=1}^{d_2} I_{\gamma_r, \gamma_s} \frac{\partial \gamma_r}{\partial \psi_i} = - I_{\psi_i, \gamma_s}, \quad i = 1, \ldots, d_1, \quad s = 1, \ldots, d_2.
    \end{equation}
\end{theorem}

The proof of Theorem~\ref{thm:1} is a direct adaptation of the argument of~\cite{Cox1987Parameter} that led to their Equation (4). To begin with, the 
log-likelihood function $l(\gbf{\psi}, \gbf{\gamma}; \gbf{X}_{1:n})$ in the $(\gbf{\psi}, \gbf{\gamma})$ parametrization and the log-likelihood function $\bar{l}(\gbf{\psi},\gbf{\lambda}; \gbf{X}_{1:n})$ in the $(\gbf{\psi}, \gbf{\lambda})$ parametrization satisfy
\begin{equation*}
    l(\gbf{\psi}, \gbf{\gamma}; \gbf{X}_{1:n}) = l(\gbf{\psi}, \bar{\gbf{\gamma}}(\gbf{\psi}, \gbf{\lambda}); \gbf{X}_{1:n}) = \bar{l}(\gbf{\psi}, \gbf{\lambda}; \gbf{X}_{1:n}).
\end{equation*}
By taking derivatives, we obtain
\begin{equation*}
    \begin{aligned}
        \frac{\partial \bar{l}}{\partial \psi_i} &= \frac{\partial l}{\partial \psi_i} + \sum_{r=1}^{d_2} \frac{\partial l}{\partial \gamma_r} \frac{\partial \gamma_r}{\partial \psi_i}, \\
        \frac{\partial^2 \bar{l}}{\partial \psi_i \partial \lambda_j} &= \sum_{s=1}^{d_2} \frac{\partial^2 l}{\partial \psi_i \partial \gamma_s} \frac{\partial \gamma_s}{\partial \lambda_j} + \sum_{r=1}^{d_2} \sum_{s=1}^{d_2} \frac{\partial^2 l}{\partial \gamma_r \partial \gamma_s} \frac{\partial \gamma_s}{\partial \lambda_j} \frac{\partial \gamma_r}{\partial \psi_i} + \sum_{r=1}^{d_2} \frac{\partial l}{\partial \gamma_r} \frac{\partial^2 \gamma_r}{\partial \psi_i \partial \lambda_j}
    \end{aligned}
\end{equation*}
for any $i = 1, \ldots, d_1$ and $j = 1, \ldots, d_2$. Taking expectations on both sides, we have that the orthogonality of $\gbf{\psi}$ and $\gbf{\lambda}$ is equivalent to
\begin{equation*}
    \sum_{s=1}^{d_2} \frac{\partial \gamma_s}{\partial \lambda_j} \Big(I_{\psi_i, \gamma_s} + \sum_{r=1}^{d_2} I_{\gamma_r, \gamma_s} \frac{\partial \gamma_r}{\partial \psi_i} \Big) = 0, \quad i = 1, \ldots, d_1, \quad j = 1, \ldots, d_2.
\end{equation*}
Since the transformation from $(\gbf{\psi}, \gbf{\lambda})$ to $(\gbf{\psi}, \gbf{\gamma})$ has to be invertible, the orthogonality condition can be written as
\begin{equation*}
    \sum_{r=1}^{d_2} I_{\gamma_r, \gamma_s} \frac{\partial \gamma_r}{\partial \psi_i} = -I_{\psi_i, \gamma_s}, \quad i = 1, \ldots, d_1, \quad s = 1, \ldots, d_2,
\end{equation*}
concluding the proof.

Theorem~\ref{thm:1} generalizes the result of Equation (4) in~\cite{Cox1987Parameter} to the multi-parameter case. Specifically, on setting $d_2 = q$ and $d_1 = 1$, Equation~\eqref{eq:ortho_entry} becomes Equation~\eqref{eq:CoxReid}. More interestingly, by allowing $d_1 > 1,$ we have identified cases for which we shall show that the answer to question (viii) of~\cite{Cox1987Parameter} is positive.


\section{Examples for global parameter orthogonality}
\label{sec:ex}

Item (viii) of Section 6 of \cite{Cox1987Parameter} states that global parameter orthogonality generally does \textit{not} hold for distributions with multiple parameters of interest, due to the  partial differential equations~\eqref{eq:ortho_entry} being unsolvable. In this section, however, we list a few examples, as summarized in Table~\ref{tab:ex}, of distributions with multiple parameters of interest for which the global orthogonality \textit{does} unexpectedly exist, with details provided afterwards. We argue that the reparameterized parameters are of practical relevance. 

\begin{table}[ht]
	\centering
	\resizebox{\textwidth}{!}{
	\begin{tabular}{l|c|c|l}
		\hline
		Distribution & \begin{tabular}{l} Parameters \\ of interest \end{tabular} & \begin{tabular}{l} Nuisance \\ parameter \end{tabular} & Orthogonal reparameterization \\
		\hline
		Generalized Gamma & $a, p$ & $d$ & $\lambda = a \exp\{\varphi(d/p)/p\}$ \\
		Location-scale family & $\mu, \sigma$ & $\theta$ & $\lambda = \sigma \exp\{-F_1(\theta)\}$ \\
		Location-scale family & $\mu, \theta$ & $\sigma$ & $\lambda = \sigma \exp\{-F_2(\theta)\}$ \\
		Two samples of Gamma & $\alpha_1, \alpha_2$ & $\theta$ & $\lambda = \theta (\alpha_1 + \alpha_2)$ \\
		Two samples of inverse-Gamma & $\alpha_1, \alpha_2$ & $\beta$ & $\lambda = (\alpha_1 + \alpha_2) / \beta$ \\
		\hline
	\end{tabular}
	}
	\caption{Examples of distributions with multiple parameters of interest for which global parameter orthogonality holds. }
	\label{tab:ex}
\end{table}

\subsection{Generalized Gamma distribution}
\label{subsec:GGamma}

Let $X$ follow a generalized Gamma distribution with probability density function
$$ f(x; a, d, p) = \frac{p}{a} \Big(\frac{y}{a}\Big)^{d - 1} \frac{1}{\Gamma (d/p)} \exp\Big\{-\Big(\frac{y}{a}\Big)^p\Big\}, \quad x > 0, $$
where $a > 0$ is the scale parameter, $d, p > 0$ are the shape parameters, and $\Gamma(\cdot)$ denotes the Gamma function. Now we treat $a$ and $p$ as the parameters of interest and $d$ as the nuisance parameter. Equation~\eqref{eq:ortho_entry} gives 
$$\frac{1}{p^2} \varphi^\prime\Big(\frac{d}{p}\Big) \frac{\partial d}{\partial a} = -\frac{1}{a}, \qquad \frac{1}{p^2} \varphi^\prime\Big(\frac{d}{p}\Big) \frac{\partial d}{\partial p} = \frac{1}{p^2} \varphi \Big(\frac{d}{p}\Big) + \frac{d}{p^3} \varphi \Big(\frac{d}{p}\Big),$$
where $\varphi(\cdot) = \Gamma^\prime(\cdot) / \Gamma(\cdot)$ is the digamma function. This system of partial differential equations yields the solution 
$$h(\lambda) = \log a + \frac{1}{p} \varphi \Big(\frac{d}{p}\Big),$$
where $h(\lambda)$ is an arbitrary function of $\lambda$. By taking $h(\lambda) = \log \lambda$, we set the reparameterized $\lambda = a \exp\{\varphi(d/p)/p\}$, which makes $a$ and $p$ globally orthogonal to $\lambda$.

\subsection{Location-scale family of single-parameter distributions}
\label{subsec:LS}

Let us now consider a location-scale family of distributions with a probability density function 
$$f(x; \mu, \sigma, \theta) = \frac{1}{\sigma} g\Big(\frac{x - \mu}{\sigma} ; \theta \Big), \quad x \in \mathbb{R},$$
where $g(z; \theta)$ is a symmetric probability density function on $\mathbb{R}$ with unknown parameter $\theta$. By symmetry, we have that $I_{\mu,\sigma}=I_{\mu,\theta}=0$, and 
\begin{equation*}
    \begin{aligned}
        I_{\sigma,\theta} &= - \frac{1}{\sigma}\int_{\mathbb{R}} \frac{z}{g(z; \theta)} \frac{\partial g(z; \theta)}{\partial z} \frac{\partial g(z; \theta)}{\partial \theta} \, \mathrm{d}z =: -\frac{1}{\sigma} f_1(\theta), \\
        I_{\sigma,\sigma} &= \frac{1}{\sigma^2} \Big[\int_{\mathbb{R}} \frac{z^2}{g(z; \theta)} \Big\{\frac{\partial g(z; \theta)}{\partial z}\Big\}^2 \, \mathrm{d}z + 2 \int_{\mathbb{R}} z \frac{\partial g(z; \theta)}{\partial z} \, \mathrm{d}z + 1\Big] =: \frac{1}{\sigma^2} f_2(\theta), \\
        I_{\theta,\theta} &= \int_{\mathbb{R}} \frac{1}{g(z; \theta)} \Big\{\frac{\partial g(z; \theta)}{\partial \theta}\Big\}^2 \, \mathrm{d}z =: f_3(\theta).
    \end{aligned}
\end{equation*}
Obviously $\mu$ is orthogonal to $\sigma$ and $\theta$. Now we treat $\mu$ and $\sigma$ as the parameters of interest and $\theta$ as the nuisance parameter. Equation~\eqref{eq:ortho_entry} gives
$$f_3(\theta) \frac{\partial \theta}{\partial \mu} = 0, \qquad f_3(\theta) \frac{\partial \theta}{\partial \sigma} = \frac{1}{\sigma} f_1(\theta),$$
which yields the solution 
$$h(\lambda) = \log \sigma - \int \frac{f_3(\theta)}{f_1(\theta)} \, \mathrm{d}\theta =: \log \sigma - F_1(\theta).$$
Taking $h(\lambda) = \log \lambda$, we can set the reparameterization to be $\lambda = \sigma \exp\{-F_1(\theta)\}$, which makes $\mu$ and $\sigma$ globally orthogonal to $\lambda$. Similarly, if we treat $\mu$ and $\theta$ as the parameters of interest and $\sigma$ as the nuisance parameter, we can also derive $\lambda = \sigma \exp\{-F_2(\theta)\}$ such that it is globally orthogonal to $\mu$ and $\theta$, where $F_2(\theta) = \int \{f_1(\theta) / f_2(\theta)\} \, \mathrm{d}\theta.$

This framework is particularly useful when we are only interested in the location and scale parameters of a distribution and 
not in the shape parameter. For instance, when we wish to provide inference for the mean and variance of heavy-tailed data, a common approach is to assume a location-scale $t$-distribution and estimate the location and scale parameters, while the degree-of-freedom parameter $\nu$ that controls tail heaviness is of less interest. By reparameterization, we can construct a nuisance parameter orthogonal to $\mu$ and $\sigma$, and thus a preliminary estimation of the shape parameter has little influence on the estimation of the location and scale parameters.

\subsection{Two samples of (inverse) Gamma distributions}
\label{subsec:TwoGamma}

We turn to the case where $X_1$ and $X_2$ are independent Gamma-distributed random variables with the same scale parameter $\theta$ but different shape parameters $\alpha_1$ and $\alpha_2$, respectively. If we treat $\alpha_1$ and $\alpha_2$ as the parameters of interest and $\theta$ as the nuisance parameter, then Equation~\eqref{eq:ortho_entry} gives
$$\frac{\alpha_1 + \alpha_2}{\theta^2} \frac{\partial \theta}{\partial \alpha_1} = -\frac{1}{\theta}, \qquad \frac{\alpha_1 + \alpha_2}{\theta^2} \frac{\partial \theta}{\partial \alpha_2} = -\frac{1}{\theta},$$
indicating that $\lambda = \theta (\alpha_1 + \alpha_2)$ is the reparameterization that makes $\alpha_1$ and $\alpha_2$ globally orthogonal to $\lambda$.

Similarly, if $X_1$ and $X_2$ are independent inverse-Gamma-distributed random variables with the same scale parameter $\beta$ but different shape parameters $\alpha_1$ and $\alpha_2$, Equation~\eqref{eq:ortho_entry} gives
$$\frac{\alpha_1 + \alpha_2}{\beta^2} \frac{\partial \beta}{\partial \alpha_1} = \frac{1}{\beta}, \qquad \frac{\alpha_1 + \alpha_2}{\beta^2} \frac{\partial \beta}{\partial \alpha_2} = \frac{1}{\beta},$$
indicating that $\lambda = (\alpha_1 + \alpha_2) / \beta$ is the reparameterization that makes $\alpha_1$ and $\alpha_2$ globally orthogonal to $\lambda$.

As demonstrated in this section, global parameter orthogonality can be achieved for certain distributions with multiple parameters of interest in special cases. For other distributions, such as the Weibull, global orthogonality does not hold. Also, in the generalized Gamma distribution, global orthogonality only arises when $d$ is treated as the nuisance parameter. Therefore, local parameter orthogonality is generally more practical and useful, and we shall explain how it contributes to computational efficiency in Section~\ref{sec:Whittle}.

\section{MLEs of location-scale \texorpdfstring{$t$}{t}-distribution}
\label{sec:ls_t}

In this section, we illustrate the practical benefits of \textit{global} parameter orthogonality by focusing on the location-scale $t$-distribution. This distribution is widely used for modeling heavy-tailed data and is parameterized by location $\mu$, scale $\sigma$, and degree-of-freedom $\nu$. The probability density function is given by
$$
f(x; \mu, \sigma, \nu) = \frac{\Gamma\left(\frac{\nu + 1}{2}\right)}{\sigma \sqrt{\nu \pi} \Gamma\left(\frac{\nu}{2}\right)} \left\{1 + \frac{(x - \mu)^2}{\nu \sigma^2}\right\}^{-\frac{\nu + 1}{2}},
$$
where $\mu$ is the location parameter, $\sigma$ is the scale parameter, and $\nu$ controls the tail heaviness. Following the notations in Subsection~\ref{subsec:LS}, we have from Proposition 4 of Appendix B of~\cite{Lange1989Robust} that 
\begin{equation*}
    f_1(\nu) = \frac{2}{(\nu + 1) (\nu + 3)}, \quad f_2(\nu) = \frac{2\nu}{\nu + 3}.
\end{equation*}
Thus $F_2(\nu) = \log \{\nu / (\nu + 1)\}$, and by treating $\mu$ and $\nu$ as the parameters of interest and $\sigma$ as the nuisance parameter, reparameterization 
$$\lambda = \frac{\sigma (\nu + 1)}{\nu}$$ 
makes $\mu$, $\nu$ and $\lambda$ totally orthogonal, in the sense that the Fisher information matrix is diagonal.

Consequently, the MLE of each of $(\mu, \lambda, \nu)$ given the other two parameters varies slowly with respect to the other two parameters, leading to a natural way to estimate the three parameters: iteratively updating the estimate of one parameter while keeping the estimates of the other two parameters fixed, as is summarized in Algorithm~\ref{algo:iter}, where the log-likelihood function for a sample $\gbf{X}_{1:n} = (X_1, \ldots, X_n)^\top$ of size $n$ is given by
\begin{equation*}
    \begin{aligned}
        l(\mu, \lambda, \nu; \gbf{X}_{1:n}) &= n \log \Gamma\Big(\frac{\nu + 1}{2}\Big) - n \log \Gamma\Big(\frac{\nu}{2}\Big) - \frac{3n}{2} \log \nu - n \log \lambda + n \log (\nu + 1) \\
        &\quad - \frac{\nu + 1}{2} \sum_{i=1}^{n} \log \Big\{1 + \frac{(\nu + 1)^2 (X_i - \mu)^2}{\nu^3 \lambda^2}\Big\} - \frac{n}{2} \log \pi.
    \end{aligned}
\end{equation*}

\begin{algorithm}[H]
\caption{Iterative MLE for location-scale $t$-distribution}
\label{algo:iter}
\begin{algorithmic}[1]
    \State Initialize $\mu^{(0)}$, $\lambda^{(0)}$, and $\nu^{(0)}$.
    \For{$k = 0, 1, 2, \ldots$ until convergence}
        \State Update $\mu^{(k+1)}$ by maximizing the log-likelihood with $\lambda = \lambda^{(k)}$, $\nu = \nu^{(k)}$ fixed.
        \State Update $\lambda^{(k+1)}$ by maximizing the log-likelihood with $\mu = \mu^{(k+1)}$, $\nu = \nu^{(k)}$ fixed.
        \State Update $\nu^{(k+1)}$ by maximizing the log-likelihood with $\mu = \mu^{(k+1)}$, $\lambda = \lambda^{(k+1)}$ fixed.
    \EndFor
\end{algorithmic}
\end{algorithm}

As has been experienced in~\cite{Lange1989Robust}, multivariate optimization methods, including BFGS~\citep{Broyden1970BFGS,Fletcher1970BFGS,Goldfarb1970BFGS,Shanno1970BFGS} algorithm (a quasi-Newton method) and Fisher's scoring algorithm, can be applied to obtain the MLEs. To motivate the benefits of parameter orthogonality, we compare the performance of the following three methods in two different parameterizations ($(\mu, \sigma, \nu)$ and $(\mu, \lambda, \nu)$) of the location-scale $t$-distribution for obtaining the MLEs of $(\mu, \sigma, \nu)$:
\begin{itemize}
    \item Iterative update as in Algorithm~\ref{algo:iter} with golden-section search for one-dimensional maximization;
    \item BFGS algorithm with variable step length;
    \item Fisher's scoring algorithm with variable step length.
\end{itemize}

It should be noted that in the orthogonal parameterization $(\mu, \lambda, \nu)$, the Fisher information matrix is diagonal, and Fisher's scoring algorithm essentially reduces to iteratively updating each parameter using a single gradient ascent step per iteration, which makes it even simpler than Algorithm~\ref{algo:iter}. EM algorithm is also considered in~\cite{Lange1989Robust}, but is not included in our comparison since simulation results indicate that it is much slower and less accurate than the three methods considered here.

Tables~\ref{tab:nu_0.5_mean} and~\ref{tab:nu_0.5_quantile} summarize the simulation results for MLE of the location-scale $t$-distribution with true values $(\mu, \sigma, \nu) = (0, 1, 0.5)$ over 10,000 replications, with results in the unshaded rows are obtained in the original $(\mu, \sigma, \nu)$ parameterization and in the shaded rows are obtained in the orthogonal $(\mu, \lambda, \nu)$ parameterization. The initial values for $(\mu, \sigma, \nu)$ are set to be the sample median, sample standard deviation, and 4, respectively. The convergence criterion is set to be the same across all methods, and $\nu$ is forced in the range $[0.1, 30]$. For each method and sample size ($n=100$ and $n=500$), we report summary statistics for time consumption (in milliseconds), number of iterations until convergence, maximized log-likelihood value, and the MLEs of $(\mu, \sigma, \nu)$. Table~\ref{tab:nu_0.5_mean} reports the average values across the 10,000 replications with standard deviations in parentheses, while Table~\ref{tab:nu_0.5_quantile} reports the $2.5\%$, $50\%$ (median), and $97.5\%$ quantiles.

\begin{table}[ht]
	\centering
	\resizebox{\textwidth}{!}{
	\begin{tabular}{lfhhhhhh}
	\hline
	\rowcolor{white}
	$n$ & Method & Time (ms) & Iterations & Log-likelihood & $\hat{\mu}$ & $\hat{\sigma}$ & $\hat{\nu}$ \\
	\hline
	\rowcolor{white}
	 & Iterative & $2.947$ ($8.516$) & $18.9$ ($51.34$) & $-3.082$ & $0.001$ ($0.157$) & $1.008$ ($0.219$) & $0.510$ ($0.075$) \\
	 & Iterative & $1.734$ ($7.461$) & $10.8$ ($45.61$) & $-3.082$ & $0.001$ ($0.157$) & $1.008$ ($0.219$) & $0.510$ ($0.075$) \\
	\rowcolor{white}
	 & BFGS & $1.003$ ($0.889$) & $18.9$ ($13.27$) & $-4.084$ & 1,683 (368,839) & 5,618,624 (369,101,504) & $1.847$ ($2.794$) \\
	 & BFGS & $0.800$ ($0.539$) & $12.7$ ($9.79$) & $-4.398$ & $-$1,981 (121,753) & 656,914 (41,945,503) & $0.574$ ($0.844$) \\
	\rowcolor{white}
	 & Fisher's Scoring & $0.764$ ($0.963$) & $19.1$ ($4.37$) & $-3.097$ & $0.001$ ($0.157$) & 5,601,792 (369,098,207) & $0.513$ ($0.150$) \\
	\multirow{-6}{*}{$100$} & Fisher's Scoring & $0.642$ ($1.119$) & $19.7$ ($4.24$) & $-3.082$ & $0.001$ ($0.157$) & $1.008$ ($0.219$) & $0.509$ ($0.075$) \\
	\hline
	\rowcolor{white}
	 & Iterative & $5.183$ ($10.786$) & $16.9$ ($36.83$) & $-3.094$ & $0.000$ ($0.069$) & $1.002$ ($0.093$) & $0.501$ ($0.031$) \\
	 & Iterative & $2.463$ ($4.396$) & $7.8$ ($14.07$) & $-3.094$ & $0.000$ ($0.069$) & $1.002$ ($0.093$) & $0.501$ ($0.031$) \\
	\rowcolor{white}
	 & BFGS & $1.645$ ($1.189$) & $20.5$ ($15.89$) & $-4.218$ & $-512.4$ (56,763) & 32,743,538 (2,005,579,849) & $0.694$ ($1.266$) \\
	 & BFGS & $1.050$ ($1.338$) & $7.6$ ($8.65$) & $-6.315$ & $-164.9$ (18,177) & 3,895,803 (227,957,061) & $0.319$ ($0.510$) \\
	\rowcolor{white}
	 & Fisher's Scoring & $1.060$ ($0.721$) & $17.6$ ($3.58$) & $-3.115$ & $0.000$ ($0.069$) & 32,467,381 (2,005,398,651) & $0.510$ ($0.277$) \\
	\multirow{-6}{*}{$500$} & Fisher's Scoring & $0.980$ ($0.526$) & $19.5$ ($3.93$) & $-3.094$ & $0.000$ ($0.069$) & $1.002$ ($0.093$) & $0.501$ ($0.031$) \\
	\hline
	\end{tabular}
	}
	\vspace{-0.2cm}
	\caption{Means and standard deviations (in parentheses) with true values $(\mu, \sigma, \nu) = (0, 1, 0.5)$ for MLEs of location-scale $t$-distribution over 10,000 replications.}
	\label{tab:nu_0.5_mean}
\end{table}

\begin{table}[ht]
	\centering
	\resizebox{\textwidth}{!}{
	\begin{tabular}{lfhhhhhh}
	\hline
	\rowcolor{white}
	$n$ & Method & Time (ms) & Iterations & Log-likelihood & $\hat{\mu}$ & $\hat{\sigma}$ & $\hat{\nu}$ \\
	\hline
	\rowcolor{white}
	 & Iterative & ($1.880$, $2.403$, $4.178$) & ($13$, $16$, $21$) & ($-3.669$, $-3.075$, $-2.530$) & ($-0.305$, $0.001$, $0.312$) & ($0.642$, $0.984$, $1.501$) & ($0.388$, $0.502$, $0.681$) \\
	 & Iterative & ($0.926$, $1.326$, $2.147$) & ($6$, $8$, $12$) & ($-3.669$, $-3.075$, $-2.530$) & ($-0.305$, $0.001$, $0.312$) & ($0.642$, $0.984$, $1.501$) & ($0.388$, $0.502$, $0.681$) \\
	\rowcolor{white}
	 & BFGS & ($0.491$, $0.829$, $2.352$) & ($5$, $15$, $47$) & ($-7.681$, $-3.601$, $-2.609$) & ($-57.32$, $-0.002$, $61.26$) & ($0.002$, $1.237$, $311.9$) & ($0.100$, $0.585$, $10.72$) \\
	 & BFGS & ($0.305$, $0.695$, $1.702$) & ($2$, $8$, $35$) & ($-9.887$, $-3.673$, $-2.594$) & ($-228.3$, $0.001$, $275.2$) & ($0.013$, $1.709$, 4,424) & ($0.100$, $0.453$, $2.579$) \\
	\rowcolor{white}
	 & Fisher's Scoring & ($0.381$, $0.719$, $1.195$) & ($13$, $18$, $30$) & ($-3.671$, $-3.075$, $-2.530$) & ($-0.305$, $0.001$, $0.312$) & ($0.642$, $0.984$, $1.503$) & ($0.388$, $0.502$, $0.682$) \\
	\multirow{-6}{*}{$100$} & Fisher's Scoring & ($0.289$, $0.602$, $0.974$) & ($13$, $19$, $30$) & ($-3.669$, $-3.075$, $-2.530$) & ($-0.305$, $0.001$, $0.312$) & ($0.642$, $0.984$, $1.501$) & ($0.388$, $0.502$, $0.681$) \\
	\hline
	\rowcolor{white}
	 & Iterative & ($3.589$, $4.285$, $6.786$) & ($14$, $15$, $18$) & ($-3.351$, $-3.092$, $-2.844$) & ($-0.134$, $-0.000$, $0.136$) & ($0.832$, $0.997$, $1.194$) & ($0.445$, $0.500$, $0.567$) \\
	 & Iterative & ($1.664$, $2.203$, $4.149$) & ($6$, $8$, $9$) & ($-3.351$, $-3.092$, $-2.844$) & ($-0.134$, $-0.000$, $0.136$) & ($0.832$, $0.997$, $1.194$) & ($0.445$, $0.500$, $0.567$) \\
	\rowcolor{white}
	 & BFGS & ($0.657$, $1.381$, $3.803$) & ($2$, $16$, $50$) & ($-10.82$, $-3.529$, $-2.881$) & ($-120.2$, $0.001$, $95.73$) & ($0.000$, $0.993$, 5,666) & ($0.100$, $0.487$, $3.822$) \\
	 & BFGS & ($0.433$, $0.888$, $2.600$) & ($2$, $5$, $32$) & ($-12.97$, $-5.846$, $-2.964$) & (-1,817, $-0.004$, 1,746) & ($0.153$, $62.44$, 107,053) & ($0.100$, $0.108$, $1.436$) \\
	\rowcolor{white}
	 & Fisher's Scoring & ($0.533$, $1.050$, $2.595$) & ($12$, $17$, $26$) & ($-3.352$, $-3.092$, $-2.844$) & ($-0.134$, $-0.000$, $0.136$) & ($0.833$, $0.998$, $1.196$) & ($0.445$, $0.500$, $0.568$) \\
	\multirow{-6}{*}{$500$} & Fisher's Scoring & ($0.501$, $1.014$, $2.154$) & ($14$, $19$, $30$) & ($-3.351$, $-3.092$, $-2.844$) & ($-0.134$, $-0.000$, $0.136$) & ($0.832$, $0.997$, $1.194$) & ($0.445$, $0.500$, $0.567$) \\
	\hline
	\end{tabular}
	}
	\vspace{-0.2cm}
	\caption{$2.5\%$, $50\%$ (median), and $97.5\%$ quantiles with true values $(\mu, \sigma, \nu) = (0, 1, 0.5)$ for MLEs of location-scale $t$-distribution over 10,000 replications.}
	\label{tab:nu_0.5_quantile}
\end{table}

It is revealed in Tables~\ref{tab:nu_0.5_mean} and~\ref{tab:nu_0.5_quantile} that for true values $(\mu, \sigma, \nu) = (0, 1, 0.5)$, both iterative update and Fisher's scoring algorithm in the proposed $(\mu, \lambda, \nu)$ parameterization outperform the others by a big margin in terms of accuracy of the MLEs, with the Fisher's scoring method being particularly effective. Iterative update in the original parameterization also works well, but takes much more time and many more iterations to converge than that in the orthogonal parameterization. In contrast, both BFGS algorithm and Fisher's scoring algorithm in the original $(\mu, \sigma, \nu)$ parameterization produce highly unstable estimates with huge variances and smaller maximized log-likelihoods, indicating failure of convergence to the true MLEs in most replications. This is because the Fisher information matrix becomes ill-conditioned when $\nu$ is small, leading to unreliable estimates of the Newton direction in both algorithms. For all methods, orthogonal parameterization improves the computational efficiency as it reduces the computation time and number of iterations until convergence.

Tables~\ref{tab:nu_1_mean} and~\ref{tab:nu_1_quantile} present the simulation results for MLEs of the location-scale $t$-distribution with true values $(\mu, \sigma, \nu) = (0, 1, 1)$ over 10,000 replications. The findings are consistent with those for $\nu = 0.5$: the orthogonal parameterization significantly improves both accuracy and computational efficiency for iterative update and Fisher's scoring methods. Unlike the case with $\nu = 0.5$, Fisher's scoring algorithm in the original parameterization now yields stable estimates. Similar patterns are observed for other values of $\nu$ (see Appendix~\ref{appendix:simulations}).

\begin{table}[H]
	\centering
	\resizebox{\textwidth}{!}{
	\begin{tabular}{lfhhhhhh}
	\hline
	\rowcolor{white}
	$n$ & Method & Time (ms) & Iterations & Log-likelihood & $\hat{\mu}$ & $\hat{\sigma}$ & $\hat{\nu}$ \\
	\hline
	\rowcolor{white}
	 & Iterative & $2.713$ ($4.922$) & $19.4$ ($34.11$) & $-1.945$ & $0.000$ ($0.144$) & $1.007$ ($0.174$) & $1.042$ ($0.210$) \\
	 & Iterative & $1.209$ ($3.830$) & $8.2$ ($26.29$) & $-1.945$ & $0.000$ ($0.144$) & $1.007$ ($0.174$) & $1.042$ ($0.210$) \\
	\rowcolor{white}
	 & BFGS & $0.798$ ($0.288$) & $17.6$ ($6.73$) & $-2.063$ & $-0.002$ ($0.770$) & $1.451$ ($1.519$) & $1.651$ ($2.116$) \\
	 & BFGS & $0.712$ ($0.278$) & $14.5$ ($7.86$) & $-2.080$ & $0.021$ ($1.144$) & $1.820$ ($5.966$) & $2.080$ ($2.062$) \\
	\rowcolor{white}
	 & Fisher's Scoring & $0.611$ ($0.208$) & $15.2$ ($2.94$) & $-1.945$ & $0.000$ ($0.144$) & $1.007$ ($0.174$) & $1.042$ ($0.210$) \\
	\multirow{-6}{*}{$100$} & Fisher's Scoring & $0.494$ ($0.374$) & $15.7$ ($2.87$) & $-1.945$ & $0.000$ ($0.144$) & $1.007$ ($0.174$) & $1.042$ ($0.210$) \\
	\hline
	\rowcolor{white}
	 & Iterative & $4.845$ ($8.231$) & $18.4$ ($31.09$) & $-1.956$ & $0.000$ ($0.064$) & $1.001$ ($0.076$) & $1.007$ ($0.081$) \\
	 & Iterative & $1.985$ ($7.190$) & $7.2$ ($28.10$) & $-1.956$ & $0.000$ ($0.064$) & $1.001$ ($0.076$) & $1.007$ ($0.081$) \\
	\rowcolor{white}
	 & BFGS & $1.277$ ($0.844$) & $19.1$ ($7.31$) & $-2.097$ & $-0.001$ ($0.551$) & $1.464$ ($1.085$) & $1.887$ ($2.062$) \\
	 & BFGS & $1.259$ ($0.858$) & $17.9$ ($6.85$) & $-2.094$ & $-0.007$ ($0.979$) & $2.169$ ($17.05$) & $1.179$ ($1.147$) \\
	\rowcolor{white}
	 & Fisher's Scoring & $0.857$ ($0.613$) & $13.5$ ($1.47$) & $-1.956$ & $0.000$ ($0.064$) & $1.001$ ($0.076$) & $1.007$ ($0.081$) \\
	\multirow{-6}{*}{$500$} & Fisher's Scoring & $0.770$ ($0.527$) & $14.1$ ($1.52$) & $-1.956$ & $0.000$ ($0.064$) & $1.001$ ($0.076$) & $1.007$ ($0.081$) \\
	\hline
	\end{tabular}
	}
	\vspace{-0.2cm}
	\caption{Means and standard deviations (in parentheses) with true values $(\mu, \sigma, \nu) = (0, 1, 1)$ for MLEs of location-scale $t$-distribution over 10,000 replications.}
	\label{tab:nu_1_mean}
\end{table}

\begin{table}[H]
	\centering
	\resizebox{\textwidth}{!}{
	\begin{tabular}{lfhhhhhh}
	\hline
	\rowcolor{white}
	$n$ & Method & Time (ms) & Iterations & Log-likelihood & $\hat{\mu}$ & $\hat{\sigma}$ & $\hat{\nu}$ \\
	\hline
	\rowcolor{white}
	 & Iterative & ($1.843$, $2.397$, $4.446$) & ($14$, $18$, $25$) & ($-2.302$, $-1.941$, $-1.592$) & ($-0.286$, $-0.000$, $0.284$) & ($0.703$, $0.994$, $1.383$) & ($0.737$, $1.007$, $1.536$) \\
	 & Iterative & ($0.736$, $1.037$, $1.853$) & ($5$, $7$, $11$) & ($-2.302$, $-1.941$, $-1.592$) & ($-0.286$, $-0.000$, $0.284$) & ($0.703$, $0.994$, $1.383$) & ($0.737$, $1.007$, $1.536$) \\
	\rowcolor{white}
	 & BFGS & ($0.424$, $0.761$, $1.296$) & ($5$, $17$, $30$) & ($-2.956$, $-1.991$, $-1.597$) & ($-1.250$, $-0.001$, $1.164$) & ($0.645$, $1.054$, $5.131$) & ($0.446$, $1.071$, $7.956$) \\
	 & BFGS & ($0.312$, $0.683$, $1.204$) & ($3$, $17$, $25$) & ($-2.828$, $-2.032$, $-1.620$) & ($-0.953$, $0.001$, $1.043$) & ($0.677$, $1.085$, $4.766$) & ($0.318$, $1.047$, $6.489$) \\
	\rowcolor{white}
	 & Fisher's Scoring & ($0.332$, $0.623$, $0.932$) & ($10$, $15$, $22$) & ($-2.302$, $-1.941$, $-1.592$) & ($-0.286$, $-0.000$, $0.284$) & ($0.703$, $0.994$, $1.383$) & ($0.737$, $1.007$, $1.536$) \\
	\multirow{-6}{*}{$100$} & Fisher's Scoring & ($0.248$, $0.446$, $0.767$) & ($11$, $15$, $22$) & ($-2.302$, $-1.941$, $-1.592$) & ($-0.286$, $-0.000$, $0.284$) & ($0.703$, $0.994$, $1.383$) & ($0.737$, $1.007$, $1.536$) \\
	\hline
	\rowcolor{white}
	 & Iterative & ($3.453$, $4.075$, $6.549$) & ($15$, $17$, $21$) & ($-2.113$, $-1.955$, $-1.796$) & ($-0.126$, $0.000$, $0.125$) & ($0.859$, $0.998$, $1.154$) & ($0.864$, $1.001$, $1.182$) \\
	 & Iterative & ($1.221$, $1.611$, $3.470$) & ($5$, $6$, $8$) & ($-2.113$, $-1.955$, $-1.796$) & ($-0.126$, $0.000$, $0.125$) & ($0.859$, $0.998$, $1.154$) & ($0.864$, $1.001$, $1.182$) \\
	\rowcolor{white}
	 & BFGS & ($0.744$, $1.145$, $2.485$) & ($5$, $21$, $30$) & ($-2.697$, $-2.013$, $-1.810$) & ($-1.253$, $-0.000$, $1.239$) & ($0.568$, $1.031$, $4.714$) & ($0.548$, $1.042$, $8.182$) \\
	 & BFGS & ($0.520$, $1.092$, $2.654$) & ($3$, $21$, $25$) & ($-3.276$, $-1.983$, $-1.806$) & ($-1.138$, $0.001$, $0.940$) & ($0.826$, $1.019$, $6.952$) & ($0.192$, $0.993$, $5.699$) \\
	\rowcolor{white}
	 & Fisher's Scoring & ($0.458$, $0.890$, $2.249$) & ($11$, $13$, $16$) & ($-2.113$, $-1.955$, $-1.796$) & ($-0.126$, $0.000$, $0.125$) & ($0.859$, $0.998$, $1.154$) & ($0.864$, $1.001$, $1.182$) \\
	\multirow{-6}{*}{$500$} & Fisher's Scoring & ($0.387$, $0.585$, $2.141$) & ($12$, $14$, $17$) & ($-2.113$, $-1.955$, $-1.796$) & ($-0.126$, $0.000$, $0.125$) & ($0.859$, $0.998$, $1.154$) & ($0.864$, $1.001$, $1.182$) \\
	\hline
	\end{tabular}
	}
	\vspace{-0.2cm}
	\caption{$2.5\%$, $50\%$ (median), and $97.5\%$ quantiles with true values $(\mu, \sigma, \nu) = (0, 1, 1)$ for MLEs of location-scale $t$-distribution over 10,000 replications.}
	\label{tab:nu_1_quantile}
\end{table}

Overall, the simulation results demonstrate the advantages of global parameter orthogonality in improving both the accuracy and computational efficiency of MLE procedures. In particular, the orthogonal parameterization $(\mu, \lambda, \nu)$ significantly enhances the performance of iterative update and Fisher's scoring algorithm, and is thus highly recommended for practical applications involving the location-scale $t$-distribution.

\section{The Whittle algorithm}
\label{sec:Whittle}

One of the additional benefits of \textit{local} parameter orthogonality that seems to be not fully exploited in the literature is that, in suitable settings, it can lead to computationally efficient algorithms. The following showcase, the~\cite{Whittle1963Fitting} algorithm for fitting multivariate autoregressive models, has implications in block Toeplitz matrix inversion~\citep{Kutikov1967Inverse,Akaike1973Block}, which is a fundamental problem in numerical analysis with wide applications in time series analysis~\citep{Whittle1963Fitting}, filtering theory~\citep{Kailath1974View}, signal processing~\citep{Wiggins1965Recursive}, control theory~\citep{Kailath1980Linear} and many other fields.


The Whittle algorithm effectively solves the following two linear systems involving a block Toeplitz matrix formed by the autocovariance matrices $\{\gbf{\Gamma}_k\}$ of a wide-sense stationary $m$-variate time series:
\begin{equation*}
	\begin{aligned}
		(\gbf{\Phi}_{1} \,\,\vdots\,\, \gbf{\Phi}_{2}\,\,\vdots\,\,  \cdots \,\,\vdots\,\, \gbf{\Phi}_{p})
	\begin{pmatrix}
	\gbf{\Gamma}_0 & \gbf{\Gamma}_1 & \cdots & \gbf{\Gamma}_{p-1} \\
	\gbf{\Gamma}_{-1} & \gbf{\Gamma}_0 & \cdots & \gbf{\Gamma}_{p-2} \\
	\vdots & \vdots & \ddots & \vdots \\
	\gbf{\Gamma}_{-p+1} & \gbf{\Gamma}_{-p+2} & \cdots & \gbf{\Gamma}_0
	\end{pmatrix} &= - (
	\gbf{\Gamma}_1 \,\,\vdots\,\, \gbf{\Gamma}_2 \,\,\vdots\,\, \cdots \,\,\vdots\,\, \gbf{\Gamma}_p), \\
	(\tilde{\gbf{\Phi}}_{p} \,\,\vdots\,\, \tilde{\gbf{\Phi}}_{p-1} \,\,\vdots\,\, \cdots \,\,\vdots\,\, \tilde{\gbf{\Phi}}_{1}) 
	\begin{pmatrix}
	\gbf{\Gamma}_0 & \gbf{\Gamma}_{1} & \cdots & \gbf{\Gamma}_{p-1} \\
	\gbf{\Gamma}_{-1} & \gbf{\Gamma}_0 & \cdots & \gbf{\Gamma}_{p-2} \\
	\vdots & \vdots & \ddots & \vdots \\
	\gbf{\Gamma}_{-p+1} & \gbf{\Gamma}_{-p+2} & \cdots & \gbf{\Gamma}_0
	\end{pmatrix} &= - (\gbf{\Gamma}_{-p} \,\,\vdots\,\, \gbf{\Gamma}_{-p+1} \,\,\vdots\,\, \cdots \,\,\vdots\,\, \gbf{\Gamma}_{-1} ),
	\end{aligned}
\end{equation*}
where the $pm \times pm$ block Toeplitz matrix on the left-hand side is denoted by $\gbf{T}$. With slight adaptations~\citep{Akaike1973Block}, it leads to an efficient inversion algorithm for $\gbf{T}$ with time complexity $O(p^2 m^3)$ instead of the usual $O(p^3 m^3)$.

We first explain how local parameter orthogonality is involved in the derivation of the Whittle algorithm. Consider a stationary zero-mean $m$-variate autoregressive model of order $p+1$, or a VAR$(p+1)$ for short:
\begin{equation}
    \label{eq:VAR(p+1)}
    \gbf{X}_t + \gbf{\Phi}_{p+1, 1} \gbf{X}_{t-1} + \cdots + \gbf{\Phi}_{p+1, p} \gbf{X}_{t-p} + \gbf{\Phi}_{p+1, p+1} \gbf{X}_{t-p-1} = \gbf{\varepsilon}_t,
\end{equation}
where $\gbf{\varepsilon}_t$'s are independent and identically distributed (i.i.d.) $m$-vectors  with probability density function $f_{\gbf{\varepsilon}}(\cdot)$. We assume that the second-order moment of $\gbf{\varepsilon}_t$ exists, and
\begin{equation*}
    \gbf{J}_{\gbf{\varepsilon}} = \E \Big[ \Big\{\frac{\partial \log f_{\gbf{\varepsilon}}(\gbf{\varepsilon})}{\partial \gbf{\varepsilon}} \Big\} \Big\{\frac{\partial \log f_{\gbf{\varepsilon}}(\gbf{\varepsilon})}{\partial \gbf{\varepsilon}} \Big\}^\top \Big]
\end{equation*}
exists, both of which may be easily verified if $\gbf{\varepsilon}_t$ follows a multivariate normal distribution or a multivariate $t$-distribution with more than 2 degrees of freedom. Let $\mathcal{F}_t=\sigma(\{\gbf{\varepsilon}_{t-s}: s \geq 0\})$ be the filtration up to time $t$. Denote by 
\begin{equation*}
\gbf{\Phi}_{p+1} = 
    (\gbf{\Phi}_{p+1, 1}\,\,\vdots\,\,  \cdots \,\,\vdots\,\, \gbf{\Phi}_{p+1, p} \,\,\vdots\,\, \gbf{\Phi}_{p+1, p+1})
\end{equation*}
the $m\times m(p+1)$ parameter matrix of the VAR$(p+1)$ model. 

Let us treat $\gbf{\Phi}_{p+1,\, p+1}$ as the parameter of interest and $\gbf{\Phi}_{p+1, s}, \quad s = 1, \ldots, p,$ as the nuisance parameters. 
Suppose that we wish to reparameterize $\gbf{\Phi}_{p+1}$ to $(\gbf{\Phi}_{p}, \gbf{\Phi}_{p+1,\, p+1})$ by transformations $\bar{\gbf{\Phi}}_{p+1, s}$:
$$\gbf{\Phi}_{p+1, s} = \bar{\gbf{\Phi}}_{p+1, s}(\gbf{\Phi}_{p}, \gbf{\Phi}_{p+1, p+1}), \quad s = 1, \ldots, p$$ 
such that the parameters $\gbf{\Phi}_{p}$ and $\gbf{\Phi}_{p+1, p+1}$ are locally orthogonal. From equation~\eqref{eq:ortho_entry},  we require that
\begin{equation*}
    \begin{aligned}
        &\sum_{r=1}^{p} \sum_{k=1}^{m} \sum_{\ell=1}^{m} I_{p+1, r, (k, \ell), s, (i, j)}(\gbf{\Phi}_{p+1}) \frac{\partial \Phi_{p+1, r, (k, \ell)}}{\partial \Phi_{p+1, p+1, (u, v)}} = - I_{p+1, p+1, (u, v), s,(i, j)}(\gbf{\Phi}_{p+1}), \\
        & \qquad \qquad s = 1, \ldots, p, \quad i, j, u, v = 1, \ldots, m,   
    \end{aligned}
\end{equation*}
where 
\begin{equation*}
    \begin{aligned}
        I_{p+1, r, (k, \ell), s, (i, j)}(\gbf{\Phi}_{p+1}) &= \E \Big\{ \frac{\partial \log f_{\gbf{\varepsilon}}(\gbf{\varepsilon}_t)}{\partial \Phi_{p+1, r, (k, \ell)}} \frac{\partial \log f_{\gbf{\varepsilon}}(\gbf{\varepsilon}_t)}{\partial \Phi_{p+1, s, (i, j)}} \Big\} \\
        &= \E \Big\{ \frac{\partial \log f_{\gbf{\varepsilon}}(\gbf{\varepsilon}_t)}{\partial \varepsilon_{t,k}}\frac{\partial \varepsilon_{t,k}}{\partial \Phi_{p+1, r, (k, \ell)}} \frac{\partial \log f_{\gbf{\varepsilon}}(\gbf{\varepsilon}_t)}{\partial \varepsilon_{t,i}} \frac{\partial \varepsilon_{t,i}}{\partial \Phi_{p+1, s, (i, j)}} \Big\} \\
        &= \E \Big\{\frac{\partial \log f_{\gbf{\varepsilon}}(\gbf{\varepsilon}_t)}{\partial \varepsilon_{t,k}} \frac{\partial \log f_{\gbf{\varepsilon}}(\gbf{\varepsilon}_t)}{\partial \varepsilon_{t,i}} X_{t-r, \ell} X_{t-s, j}\Big\} \\
        &= \E \Big[ X_{t-r, \ell} X_{t-s, j} \E \Big\{\frac{\partial \log f_{\gbf{\varepsilon}}(\gbf{\varepsilon}_t)}{\partial \varepsilon_{t,k}} \frac{\partial \log f_{\gbf{\varepsilon}}(\gbf{\varepsilon}_t)}{\partial \varepsilon_{t,i}}  \Big| \mathcal{F}_{t-1} \Big\} \Big]\\
        &= \E ( X_{t-r, \ell} X_{t-s, j} ) \E \Big\{\frac{\partial \log f_{\gbf{\varepsilon}}(\gbf{\varepsilon}_t)}{\partial \varepsilon_{t,k}} \frac{\partial \log f_{\gbf{\varepsilon}}(\gbf{\varepsilon}_t)}{\partial \varepsilon_{t,i}}\Big\} \\
        &= \Gamma_{s-r, (\ell, j)} J_{\gbf{\varepsilon}, (k,i)}.
    \end{aligned}
\end{equation*}
Hence 
\begin{equation}
    \label{eq:ortho_VAR_entry}
    \sum_{r=1}^{p} \sum_{k=1}^{m} \sum_{\ell=1}^{m} J_{\gbf{\varepsilon}, (k,i)} \Gamma_{s-r, (\ell, j)} \frac{\partial \Phi_{p+1, r, (k, \ell)}}{\partial \Phi_{p+1, p+1, (u, v)}} = - J_{\gbf{\varepsilon}, (u,i)} \Gamma_{s-p-1, (v, j)}
\end{equation}
has to hold for all $s = 1, \ldots, p$ and $i, j, u, v = 1, \ldots, m$.

Now for a backward VAR$(p)$ modeling scheme with parameters $\widetilde{\gbf{\Phi}}_p$, we have
\begin{equation}
    \label{eq:b_VAR(p)}
    \gbf{X}_t + \widetilde{\gbf{\Phi}}_{p, 1} \gbf{X}_{t+1} + \cdots + \widetilde{\gbf{\Phi}}_{p, p} \gbf{X}_{t+p} = \widetilde{\gbf{\eta}}_t,
\end{equation}
where the $\widetilde{\gbf{\eta}}_t$'s are i.i.d. with probability density function $f_{\gbf{\eta}}(\cdot)$. By the Yule--Walker equations,
\begin{equation}
    \label{eq:Yule_Walker}
    \sum_{r=1}^{p} \widetilde{\gbf{\Phi}}_{p, p+1-r} \gbf{\Gamma}_{s-r} = \sum_{r=1}^{p} \widetilde{\gbf{\Phi}}_{p, r} \gbf{\Gamma}_{r-p-1+s} = - \gbf{\Gamma}_{s-p-1}, \quad s = 1, \ldots, p,
\end{equation}
which can be written entrywise as
\begin{equation}
    \label{eq:Yule_Walker_entry}
    \sum_{r=1}^{p} \sum_{\ell=1}^{m} \widetilde{\Phi}_{p, p+1-r, (i, \ell)} \Gamma_{s-r, (\ell, j)} = - \Gamma_{s-p-1, (i, j)}, \quad s = 1, \ldots, p, \quad i, j = 1, \ldots, m.
\end{equation}

Equations~\eqref{eq:ortho_VAR_entry} and~\eqref{eq:Yule_Walker_entry} together yield a local orthogonal reparameterization of the VAR$(p+1)$ model parameters~\eqref{eq:VAR(p+1)} in terms of the VAR$(p)$ model parameters~\eqref{eq:b_VAR(p)}. Specifically, we may set 
\begin{equation}
    \label{eq:reparam}
    \gbf{\Phi}_{p+1, r} = \gbf{\Phi}_{p, r} + \gbf{\Phi}_{p+1, p+1} \widetilde{\gbf{\Phi}}_{p, p+1-r}, \quad r = 1, \ldots, p,
\end{equation}
which are identical to the forward recursion equations of the Whittle algorithm. Then
\begin{equation}
    \label{eq:reparam_entry}
    \frac{\partial \Phi_{p+1, r, (k, \ell)}}{\partial \Phi_{p+1, p+1, (u, v)}} = \widetilde{\Phi}_{p, p+1-r, (v, \ell)}\, \delta_{u, k}.
\end{equation}
Together with~\eqref{eq:Yule_Walker_entry},~\eqref{eq:reparam_entry} secures the orthogonality condition~\eqref{eq:ortho_VAR_entry}. Therefore in the reparameterization~\eqref{eq:reparam}, $\gbf{\Phi}_{p}$ and $\gbf{\Phi}_{p+1, p+1}$ are locally orthogonal, and their Whittle estimates are asymptotically independent.

In fact, assuming that transformations $\bar{\gbf{\Phi}}_{p+1, s}$,  $s = 1, \ldots, p$, are linear, Whittle's forward recursion equations are the only possible transformation that ensures the local orthogonality condition~\eqref{eq:ortho_VAR_entry}. This is because~\eqref{eq:ortho_VAR_entry} in matrix form is equivalent to 
\begin{equation*}
    \begin{aligned}
		&\ \gbf{J}_{\gbf{\varepsilon}} ~
    	\Big( \frac{\partial \gbf{\Phi}_{p+1, 1}}{\partial \Phi_{p+1, p+1, (u, v)}} ~ \cdots ~ \frac{\partial \gbf{\Phi}_{p+1, p}}{\partial \Phi_{p+1, p+1, (u, v)}} \Big)
    	\begin{pmatrix}
    	    \gbf{\Gamma}_{0} & \cdots & \gbf{\Gamma}_{p-1} \\
    	    \vdots & \ddots & \vdots \\
    	    \gbf{\Gamma}_{-p+1} & \cdots & \gbf{\Gamma}_{0}
    	\end{pmatrix} \\
    	=&- \gbf{J}_{\gbf{\varepsilon}} \gbf{E}_{u, v} ~
    	(\gbf{\Gamma}_{-p} ~ \cdots ~ \gbf{\Gamma}_{-1}),
	\end{aligned}
\end{equation*}
and thus 
\begin{equation}
    \label{eq:ortho_VAR_mat}
	\begin{aligned}
		&\ \Big( \frac{\partial \gbf{\Phi}_{p+1, 1}}{\partial \Phi_{p+1, p+1, (u, v)}} ~ \cdots ~ \frac{\partial \gbf{\Phi}_{p+1, p}}{\partial \Phi_{p+1, p+1, (u, v)}} \Big) \\
		=& - \gbf{E}_{u, v} ~
    	(\gbf{\Gamma}_{-p} ~ \cdots ~ \gbf{\Gamma}_{-1})
		\begin{pmatrix}
			\gbf{\Gamma}_{0} & \cdots & \gbf{\Gamma}_{p-1} \\
			\vdots & \ddots & \vdots \\
			\gbf{\Gamma}_{-p+1} & \cdots & \gbf{\Gamma}_{0}
		\end{pmatrix}^{-1} ,
	\end{aligned}
\end{equation}
where the $(k, \ell)$-th entry of $\partial \gbf{\Phi}_{p+1, r} / \partial \Phi_{p+1, p+1, (u, v)} \in \mathbb{R}^{m \times m} $  is  $\partial \Phi_{p+1, r, (k, \ell)} / \partial \Phi_{p+1, p+1, (u, v)}$. Meanwhile, equation~\eqref{eq:Yule_Walker} implies that
\begin{equation*}
    (\widetilde{\gbf{\Phi}}_{p, p} ~ \cdots ~ \widetilde{\gbf{\Phi}}_{p, 1})
    \begin{pmatrix}
        \gbf{\Gamma}_{0} & \cdots & \gbf{\Gamma}_{p-1} \\
        \vdots & \ddots & \vdots \\
        \gbf{\Gamma}_{-p+1} & \cdots & \gbf{\Gamma}_{0}
    \end{pmatrix} 
    = - (\gbf{\Gamma}_{-p} ~ \cdots ~ \gbf{\Gamma}_{-1}),
\end{equation*}
which leads to 
\begin{equation}
    \label{eq:Yule_Walker_mat}
    (\widetilde{\gbf{\Phi}}_{p, p} ~ \cdots ~ \widetilde{\gbf{\Phi}}_{p, 1})
    = - ( \gbf{\Gamma}_{-p} ~ \cdots ~ \gbf{\Gamma}_{-1} )
    \begin{pmatrix}
        \gbf{\Gamma}_{0} & \cdots & \gbf{\Gamma}_{p-1} \\
        \vdots & \ddots & \vdots \\
        \gbf{\Gamma}_{-p+1} & \cdots & \gbf{\Gamma}_{0}
    \end{pmatrix}^{-1}.
\end{equation}
Comparing equations~\eqref{eq:ortho_VAR_mat} and~\eqref{eq:Yule_Walker_mat}, we obtain that 
\begin{equation*}
    \frac{\partial \gbf{\Phi}_{p+1, r}}{\partial \Phi_{p+1, p+1, (u, v)}} = \gbf{E}_{u, v} \widetilde{\gbf{\Phi}}_{p, p+1-r}, \quad r = 1, \ldots, p, \quad u, v = 1, \ldots, m.
\end{equation*}
Note that when all entries of $\gbf{\Phi}_{p+1, p+1}$ are 0, the VAR$(p+1)$ model reduces to a VAR$(p)$ model, and $\gbf{\Phi}_{p+1, r}=\gbf{\Phi}_{p, r}$. Thus
\begin{equation*}
    \begin{aligned}
        \gbf{\Phi}_{p+1, r} &= \gbf{\Phi}_{p, r} + \sum_{u=1}^{m} \sum_{v=1}^{m} \Phi_{p+1, p+1, (u, v)} \gbf{E}_{u, v} \widetilde{\gbf{\Phi}}_{p, p+1-r} \\
        &= \gbf{\Phi}_{p, r} + \gbf{\Phi}_{p+1, p+1} \widetilde{\gbf{\Phi}}_{p, p+1-r}, \quad r = 1, \ldots, p,
    \end{aligned}
\end{equation*}
which is exactly the forward recursion equation~\eqref{eq:reparam} of the Whittle algorithm.

Similar results for the backward recursion equation of~\cite{Whittle1963Fitting}
\begin{equation*}
    \widetilde{\gbf{\Phi}}_{p+1, r} = \widetilde{\gbf{\Phi}}_{p, r} + \widetilde{\gbf{\Phi}}_{p+1, p+1} \gbf{\Phi}_{p, p+1-r}, \quad r = 1, \ldots, p,
\end{equation*}
can be obtained by applying the same arguments to the backward VAR$(p+1)$ model 
\begin{equation*}
    \gbf{X}_t + \widetilde{\gbf{\Phi}}_{p+1, 1} \gbf{X}_{t+1} + \cdots + \widetilde{\gbf{\Phi}}_{p+1, p} \gbf{X}_{t+p} + \widetilde{\gbf{\Phi}}_{p+1, p+1} \gbf{X}_{t+p+1} = \widetilde{\gbf{\varepsilon}}_t
\end{equation*}
and the forward VAR$(p)$ model
\begin{equation*}
    \gbf{X}_t + \gbf{\Phi}_{p, 1} \gbf{X}_{t-1} + \cdots + \gbf{\Phi}_{p, p} \gbf{X}_{t-p} = \gbf{\eta}_t.
\end{equation*}

It is also revealed that Whittle's estimate for the partial autocorrelation matrix $\gbf{\Phi}_{p+1, p+1}$ is locally orthogonal to the estimates of the parameters $\gbf{\Phi}_{p}$ of the VAR$(p)$ model, and thus they are asymptotically independent. 

\section{Concluding remarks}
\label{sec:conclusion}

This note addresses some of the issues surrounding parameter orthogonality left unexamined in~\citet{Cox1987Parameter}, and 
not yet been fully explored in the literature. We extended the parameter orthogonality condition to cases with multiple parameters of interest, and provided several concrete examples to illustrate the possibilities of global orthogonal parameters. Furthermore, we applied the parameter orthogonality concept to improve the MLEs of the location-scale $t$-distribution and demonstrated that the global orthogonal parameterization $(\mu, \sigma(\nu + 1) / \nu, \nu)$ outperforms the original one in terms of both estimation accuracy and computational efficiency through simulation studies. We also showed how local parameter orthogonality plays a crucial role in the derivation of the Whittle algorithm for fitting multivariate autoregressive models, which has important implications in fast inversion of a block Toeplitz matrix.

It should be further pointed out that the applications of both global and local parameter orthogonality in efficient algorithm design are not limited to the examples presented in this note. For instance, distributions other than the location-scale $t$-distribution, may also benefit from reparameterization based on global parameter orthogonality in MLE procedures. Such distributions may be applied to model the error terms in linear and nonlinear regression models for robust estimation, and many other statistical modeling scenarios. Also, for the (block) Hankel matrix, a counterpart of the Toeplitz matrix reflected about the skew-diagonal, a fast inversion algorithm may be developed based on local parameter orthogonality in parallel with the Whittle algorithm, as discussed analogously in~\citet{Rissanen1973Algorithms}. These topics deserve further investigation in future research.

\appendix
\section{Additional simulation results}
\label{appendix:simulations}

Below are additional simulation results for the MLEs of the location-scale $t$-distribution with true values $(\mu, \sigma, \nu) = (0, 1, 2)$ and $(0, 1, 4)$.

\begin{table}[H]
	\centering
	\resizebox{\textwidth}{!}{
	\begin{tabular}{lfhhhhhh}
	\hline
	\rowcolor{white}
	$n$ & Method & Time (ms) & Iterations & Log-likelihood & $\hat{\mu}$ & $\hat{\sigma}$ & $\hat{\nu}$ \\
	\hline
	\rowcolor{white}
	 & Iterative & $2.893$ ($4.538$) & $22.4$ ($32.73$) & $-1.375$ & $-0.001$ ($0.130$) & $1.009$ ($0.147$) & $2.277$ ($1.226$) \\
	 & Iterative & $1.047$ ($3.373$) & $7.6$ ($24.36$) & $-1.375$ & $-0.001$ ($0.130$) & $1.009$ ($0.147$) & $2.277$ ($1.226$) \\
	\rowcolor{white}
	 & BFGS & $0.617$ ($0.198$) & $13.7$ ($2.63$) & $-1.378$ & $-0.002$ ($0.142$) & $1.017$ ($0.205$) & $2.313$ ($1.408$) \\
	 & BFGS & $0.546$ ($0.186$) & $12.1$ ($3.62$) & $-1.394$ & $-0.001$ ($0.150$) & $1.091$ ($0.258$) & $2.800$ ($1.763$) \\
	\rowcolor{white}
	 & Fisher's Scoring & $0.586$ ($0.186$) & $13.5$ ($3.16$) & $-1.375$ & $-0.001$ ($0.130$) & $1.009$ ($0.147$) & $2.277$ ($1.226$) \\
	\multirow{-6}{*}{$100$} & Fisher's Scoring & $0.438$ ($0.182$) & $13.6$ ($3.23$) & $-1.375$ & $-0.001$ ($0.130$) & $1.009$ ($0.147$) & $2.277$ ($1.226$) \\
	\hline
	\rowcolor{white}
	 & Iterative & $4.965$ ($5.521$) & $20.5$ ($19.70$) & $-1.386$ & $-0.000$ ($0.058$) & $1.003$ ($0.064$) & $2.037$ ($0.245$) \\
	 & Iterative & $1.584$ ($3.882$) & $6.3$ ($14.12$) & $-1.386$ & $-0.000$ ($0.058$) & $1.003$ ($0.064$) & $2.037$ ($0.245$) \\
	\rowcolor{white}
	 & BFGS & $0.974$ ($0.941$) & $13.9$ ($2.56$) & $-1.390$ & $0.000$ ($0.081$) & $1.012$ ($0.138$) & $2.108$ ($1.072$) \\
	 & BFGS & $0.857$ ($0.489$) & $12.8$ ($3.29$) & $-1.395$ & $-0.000$ ($0.063$) & $1.044$ ($0.164$) & $2.402$ ($1.304$) \\
	\rowcolor{white}
	 & Fisher's Scoring & $0.784$ ($0.482$) & $11.1$ ($1.63$) & $-1.386$ & $-0.000$ ($0.058$) & $1.003$ ($0.064$) & $2.037$ ($0.245$) \\
	\multirow{-6}{*}{$500$} & Fisher's Scoring & $0.668$ ($0.376$) & $11.1$ ($1.77$) & $-1.386$ & $-0.000$ ($0.058$) & $1.003$ ($0.064$) & $2.037$ ($0.245$) \\
	\hline
	\end{tabular}
	}
	\vspace{-0.2cm}
	\caption{Means and standard deviations (in parentheses) with true values $(\mu, \sigma, \nu) = (0, 1, 2)$ for MLEs of location-scale $t$-distribution over 10,000 replications.}
	\label{tab:nu_2_mean}
\end{table}

\begin{table}[H]
	\centering
	\resizebox{\textwidth}{!}{
	\begin{tabular}{lfhhhhhh}
	\hline
	\rowcolor{white}
	$n$ & Method & Time (ms) & Iterations & Log-likelihood & $\hat{\mu}$ & $\hat{\sigma}$ & $\hat{\nu}$ \\
	\hline
	\rowcolor{white}
	 & Iterative & ($1.876$, $2.594$, $4.763$) & ($15$, $21$, $32$) & ($-1.621$, $-1.374$, $-1.135$) & ($-0.254$, $-0.003$, $0.251$) & ($0.752$, $1.000$, $1.327$) & ($1.333$, $2.043$, $4.433$) \\
	 & Iterative & ($0.647$, $0.915$, $1.584$) & ($5$, $7$, $10$) & ($-1.621$, $-1.374$, $-1.135$) & ($-0.254$, $-0.003$, $0.251$) & ($0.752$, $1.000$, $1.327$) & ($1.333$, $2.043$, $4.433$) \\
	\rowcolor{white}
	 & BFGS & ($0.354$, $0.600$, $0.936$) & ($10$, $13$, $20$) & ($-1.636$, $-1.375$, $-1.135$) & ($-0.263$, $-0.003$, $0.258$) & ($0.750$, $1.001$, $1.342$) & ($1.321$, $2.050$, $4.640$) \\
	 & BFGS & ($0.327$, $0.475$, $0.877$) & ($3$, $13$, $17$) & ($-1.717$, $-1.384$, $-1.136$) & ($-0.285$, $-0.002$, $0.286$) & ($0.767$, $1.035$, $1.797$) & ($1.368$, $2.210$, $6.872$) \\
	\rowcolor{white}
	 & Fisher's Scoring & ($0.293$, $0.634$, $0.896$) & ($9$, $13$, $21$) & ($-1.621$, $-1.374$, $-1.135$) & ($-0.254$, $-0.003$, $0.251$) & ($0.752$, $1.000$, $1.327$) & ($1.333$, $2.043$, $4.433$) \\
	\multirow{-6}{*}{$100$} & Fisher's Scoring & ($0.204$, $0.372$, $0.736$) & ($9$, $13$, $21$) & ($-1.621$, $-1.374$, $-1.135$) & ($-0.254$, $-0.003$, $0.251$) & ($0.752$, $1.000$, $1.327$) & ($1.333$, $2.043$, $4.433$) \\
	\hline
	\rowcolor{white}
	 & Iterative & ($3.673$, $4.426$, $6.793$) & ($17$, $20$, $24$) & ($-1.500$, $-1.386$, $-1.275$) & ($-0.116$, $-0.000$, $0.114$) & ($0.884$, $1.001$, $1.132$) & ($1.641$, $2.007$, $2.592$) \\
	 & Iterative & ($1.127$, $1.398$, $3.048$) & ($5$, $6$, $8$) & ($-1.500$, $-1.386$, $-1.275$) & ($-0.116$, $-0.000$, $0.114$) & ($0.884$, $1.001$, $1.132$) & ($1.641$, $2.007$, $2.592$) \\
	\rowcolor{white}
	 & BFGS & ($0.505$, $1.016$, $2.209$) & ($10$, $14$, $20$) & ($-1.513$, $-1.387$, $-1.275$) & ($-0.119$, $-0.000$, $0.119$) & ($0.883$, $1.002$, $1.143$) & ($1.634$, $2.011$, $2.647$) \\
	 & BFGS & ($0.489$, $0.693$, $1.581$) & ($4$, $13$, $18$) & ($-1.561$, $-1.390$, $-1.276$) & ($-0.118$, $-0.000$, $0.118$) & ($0.887$, $1.010$, $1.575$) & ($1.653$, $2.047$, $7.119$) \\
	\rowcolor{white}
	 & Fisher's Scoring & ($0.375$, $0.890$, $1.123$) & ($8$, $11$, $15$) & ($-1.500$, $-1.386$, $-1.275$) & ($-0.116$, $-0.000$, $0.114$) & ($0.884$, $1.001$, $1.132$) & ($1.641$, $2.007$, $2.592$) \\
	\multirow{-6}{*}{$500$} & Fisher's Scoring & ($0.295$, $0.503$, $1.750$) & ($8$, $11$, $15$) & ($-1.500$, $-1.386$, $-1.275$) & ($-0.116$, $-0.000$, $0.114$) & ($0.884$, $1.001$, $1.132$) & ($1.641$, $2.007$, $2.592$) \\
	\hline
	\end{tabular}
	}
	\vspace{-0.2cm}
	\caption{$2.5\%$, $50\%$ (median), and $97.5\%$ quantiles with true values $(\mu, \sigma, \nu) = (0, 1, 2)$ for MLEs of location-scale $t$-distribution over 10,000 replications.}
	\label{tab:nu_2_quantile}
\end{table}

\begin{table}[H]
	\centering
	\resizebox{\textwidth}{!}{
	\begin{tabular}{lfhhhhhh}
	\hline
	\rowcolor{white}
	$n$ & Method & Time (ms) & Iterations & Log-likelihood & $\hat{\mu}$ & $\hat{\sigma}$ & $\hat{\nu}$ \\
	\hline
	\rowcolor{white}
	 & Iterative & $3.559$ ($7.109$) & $28.5$ ($54.62$) & $-1.094$ & $-0.001$ ($0.118$) & $1.013$ ($0.130$) & $6.618$ ($6.531$) \\
	 & Iterative & $1.164$ ($5.045$) & $8.6$ ($35.76$) & $-1.094$ & $-0.001$ ($0.118$) & $1.013$ ($0.130$) & $6.618$ ($6.531$) \\
	\rowcolor{white}
	 & BFGS & $0.531$ ($0.194$) & $12.0$ ($1.93$) & $-1.094$ & $-0.001$ ($0.118$) & $1.013$ ($0.130$) & $6.618$ ($6.531$) \\
	 & BFGS & $0.487$ ($0.165$) & $11.9$ ($1.66$) & $-1.095$ & $-0.001$ ($0.118$) & $1.014$ ($0.130$) & $6.626$ ($6.529$) \\
	\rowcolor{white}
	 & Fisher's Scoring & $0.580$ ($0.220$) & $13.9$ ($5.43$) & $-1.094$ & $-0.001$ ($0.118$) & $1.014$ ($0.132$) & $6.618$ ($6.531$) \\
	\multirow{-6}{*}{$100$} & Fisher's Scoring & $0.421$ ($0.213$) & $14.0$ ($5.31$) & $-1.094$ & $-0.001$ ($0.118$) & $1.013$ ($0.130$) & $6.618$ ($6.531$) \\
	\hline
	\rowcolor{white}
	 & Iterative & $5.162$ ($5.175$) & $22.1$ ($19.88$) & $-1.107$ & $-0.000$ ($0.053$) & $1.003$ ($0.057$) & $4.236$ ($0.972$) \\
	 & Iterative & $1.437$ ($3.615$) & $5.8$ ($14.09$) & $-1.107$ & $-0.000$ ($0.053$) & $1.003$ ($0.057$) & $4.236$ ($0.972$) \\
	\rowcolor{white}
	 & BFGS & $0.810$ ($0.504$) & $11.8$ ($1.57$) & $-1.107$ & $-0.000$ ($0.053$) & $1.003$ ($0.057$) & $4.236$ ($0.972$) \\
	 & BFGS & $0.749$ ($0.673$) & $11.7$ ($1.28$) & $-1.107$ & $-0.000$ ($0.053$) & $1.003$ ($0.057$) & $4.236$ ($0.972$) \\
	\rowcolor{white}
	 & Fisher's Scoring & $0.742$ ($0.580$) & $10.4$ ($1.71$) & $-1.107$ & $-0.000$ ($0.053$) & $1.003$ ($0.057$) & $4.236$ ($0.972$) \\
	\multirow{-6}{*}{$500$} & Fisher's Scoring & $0.596$ ($0.346$) & $10.5$ ($1.71$) & $-1.107$ & $-0.000$ ($0.053$) & $1.003$ ($0.057$) & $4.236$ ($0.972$) \\
	\hline
	\end{tabular}
	}
	\vspace{-0.2cm}
	\caption{Means and standard deviations (in parentheses) with true values $(\mu, \sigma, \nu) = (0, 1, 4)$ for MLEs of location-scale $t$-distribution over 10,000 replications.}
	\label{tab:nu_4_mean}
\end{table}

\begin{table}[H]
	\centering
	\resizebox{\textwidth}{!}{
	\begin{tabular}{lfhhhhhh}
	\hline
	\rowcolor{white}
	$n$ & Method & Time (ms) & Iterations & Log-likelihood & $\hat{\mu}$ & $\hat{\sigma}$ & $\hat{\nu}$ \\
	\hline
	\rowcolor{white}
	 & Iterative & ($1.148$, $2.866$, $6.197$) & ($8$, $23$, $49$) & ($-1.282$, $-1.095$, $-0.900$) & ($-0.229$, $-0.003$, $0.235$) & ($0.774$, $1.006$, $1.281$) & ($2.270$, $4.325$, $30.000$) \\
	 & Iterative & ($0.623$, $0.809$, $1.913$) & ($5$, $6$, $12$) & ($-1.282$, $-1.095$, $-0.900$) & ($-0.229$, $-0.003$, $0.235$) & ($0.774$, $1.006$, $1.281$) & ($2.270$, $4.325$, $30.000$) \\
	\rowcolor{white}
	 & BFGS & ($0.316$, $0.461$, $0.824$) & ($9$, $12$, $17$) & ($-1.282$, $-1.095$, $-0.900$) & ($-0.229$, $-0.003$, $0.235$) & ($0.774$, $1.007$, $1.282$) & ($2.270$, $4.325$, $30.000$) \\
	 & BFGS & ($0.320$, $0.420$, $0.816$) & ($9$, $12$, $16$) & ($-1.284$, $-1.095$, $-0.900$) & ($-0.229$, $-0.003$, $0.235$) & ($0.775$, $1.007$, $1.282$) & ($2.274$, $4.335$, $30.000$) \\
	\rowcolor{white}
	 & Fisher's Scoring & ($0.281$, $0.600$, $0.982$) & ($6$, $13$, $26$) & ($-1.282$, $-1.095$, $-0.901$) & ($-0.229$, $-0.003$, $0.235$) & ($0.774$, $1.007$, $1.289$) & ($2.270$, $4.325$, $30.000$) \\
	\multirow{-6}{*}{$100$} & Fisher's Scoring & ($0.188$, $0.347$, $0.771$) & ($8$, $13$, $26$) & ($-1.282$, $-1.095$, $-0.900$) & ($-0.229$, $-0.003$, $0.235$) & ($0.774$, $1.006$, $1.281$) & ($2.270$, $4.325$, $30.000$) \\
	\hline
	\rowcolor{white}
	 & Iterative & ($3.527$, $4.666$, $7.222$) & ($16$, $21$, $29$) & ($-1.191$, $-1.107$, $-1.020$) & ($-0.102$, $0.000$, $0.102$) & ($0.895$, $1.002$, $1.117$) & ($2.961$, $4.051$, $6.531$) \\
	 & Iterative & ($1.062$, $1.303$, $2.807$) & ($5$, $5$, $8$) & ($-1.191$, $-1.107$, $-1.020$) & ($-0.102$, $0.000$, $0.102$) & ($0.895$, $1.002$, $1.117$) & ($2.961$, $4.051$, $6.531$) \\
	\rowcolor{white}
	 & BFGS & ($0.421$, $0.717$, $1.998$) & ($9$, $12$, $15$) & ($-1.191$, $-1.107$, $-1.020$) & ($-0.102$, $0.000$, $0.102$) & ($0.895$, $1.002$, $1.117$) & ($2.961$, $4.051$, $6.531$) \\
	 & BFGS & ($0.436$, $0.569$, $1.284$) & ($10$, $12$, $15$) & ($-1.191$, $-1.107$, $-1.020$) & ($-0.102$, $0.000$, $0.102$) & ($0.895$, $1.002$, $1.117$) & ($2.961$, $4.051$, $6.536$) \\
	\rowcolor{white}
	 & Fisher's Scoring & ($0.344$, $0.850$, $1.103$) & ($8$, $10$, $14$) & ($-1.191$, $-1.107$, $-1.020$) & ($-0.102$, $0.000$, $0.102$) & ($0.895$, $1.002$, $1.117$) & ($2.961$, $4.051$, $6.531$) \\
	\multirow{-6}{*}{$500$} & Fisher's Scoring & ($0.278$, $0.425$, $1.058$) & ($8$, $10$, $14$) & ($-1.191$, $-1.107$, $-1.020$) & ($-0.102$, $0.000$, $0.102$) & ($0.895$, $1.002$, $1.117$) & ($2.961$, $4.051$, $6.531$) \\
	\hline
	\end{tabular}
	}
	\vspace{-0.2cm}
	\caption{$2.5\%$, $50\%$ (median), and $97.5\%$ quantiles with true values $(\mu, \sigma, \nu) = (0, 1, 4)$ for MLEs of location-scale $t$-distribution over 10,000 replications.}
	\label{tab:nu_4_quantile}
\end{table}

\bibliographystyle{dcu}
\bibliography{ref}

\end{document}